\documentclass{aa}
\usepackage{graphicx}
\usepackage{natbib}
\usepackage{txfonts}

\begin{document}

\title{Optical imaging of L723: the structure of  HH~223.~\thanks{Based on observations made with the 
2.6~m Nordic Optical Telescope
operated at the Observatorio del Roque 
de los Muchachos of the Instituto de Astrof\' \i sica de Canarias. } 
}

\author{Rosario L\'opez \inst{1}
\and Robert Estalella \inst{1}
\and Gabriel G\'omez \inst{2}
\and Angels Riera \inst{3,1}
}

\offprints{R. L\'opez}

\institute{
Departament d'Astronomia i Meteorologia, Universitat de Barcelona,
Av.\ Diagonal 647, E-08028 Barcelona, Spain;
email: rosario@am.ub.es, robert.estalella@am.ub.es
\and
Instituto de Astrof\'{\i}sica de Canarias, E38200 La Laguna, Tenerife, Spain; email:
ggv@ll.iac.es
\and
Departament de F\'\i sica i Enginyeria Nuclear, Universitat
Polit\`ecnica de Catalunya, Av. V\' {\i}ctor Balaguer s/n, E-08800
Vilanova i la Geltr\'u, Spain; email:angels.riera@upc.es
}


\titlerunning{Optical imaging of  HH~223 }

\abstract{}
{We imaged the Lynds~723 dark nebula (L723) with the aim of studying the morphology of the
Herbig-Haro object HH~223 and other line-emission nebula detected in the region.}
{We obtained deep narrow-band images 
in the H$\alpha$ and [SII] lines and in the continuum nearby H$\alpha$ 
of a field of $\sim 5'$ of the L723 dark nebula  
centered on HH~223.}
{The H$\alpha$  and [SII] images reveal the detailed morphology of HH~223, 
unresolved in previous optical images.
Both images show a quite complex knotty, wiggling 
structure embedded in a low-emission nebula. Comparison between the  
[SII] and H$\alpha$ fluxes of the knots are indicative of 
variations in the excitation conditions through HH~223.
In addition, several other 
faint nebula are detected in H$\alpha$ a few arcmin to the SE and 
to the NW of HH~223,
all of them lying projected  onto the east-west pair of lobes of the quadrupolar CO 
outflow. Comparison between the H$\alpha$ and the continuum images 
confirms the HH-like nature of the Vrba object V83, while the
Vrba objects V84 and V85 are identified as faint field stars.}{}


\maketitle

\section{Introduction}

Lynds 723 (L723) is an isolate dark cloud located at a distance of 300$\pm$150 pc 
(Goldsmith et al.~\citealp{gol84}) with clear signs of star formation activity. It is 
one of the few known cases where  a  
quadrupolar CO outflow has been found. The outflow is formed by two separate pairs of
bipolar CO lobes 
(a larger pair, of  $\sim 7'$ long, lying in an east-west
orientation and another smaller pair, of $\sim 4'$ long,
lying in a north-south orientation). 
The four CO lobes emanate from  a common center, 
where  
the Class 0 source IRAS 19156+1906 is located. Since the discovery of the
outflow, several scenarios have been proposed to explain its 
peculiar morphology
(Goldsmith et al.\  \citealp{gol84}; Avery et al. \citealp{ave90}; 
Lee et al.\ \citealp{lee02} and references therein).  

Anglada et al.\ \cite{ang91} discovered two 3.6~cm radio continuum sources, 
VLA1 and VLA2, towards the center of the quadrupolar outflow, 
both within the error 
ellipsoide of the IRAS source. Higher  resolution observations by Anglada et
al.\ \cite{ang96} found that the VLA2 source appears elongated approximately 
along the axis direction of the east-west CO bipolar outflow. Their results suggest 
that this source  is a thermal radio jet, being related to the 
excitation of the east-west CO outflow. 
Later observations led to conclude that only VLA2 is a young stellar object
(YSO), being  
associated with the  CO outflow, while VLA1 is most likely a 
field source.
In fact, only VLA2 is associated with circumstellar gas traced by 
millimeter continuum emission (Cabrit \& Andr\'e  \citealp{cab91}; Reipurth  
et al.\ \citealp{rei93}), and in addition, VLA2 is embedded in  high density
molecular gas traced by ammonia  
(Girart et al.\ \citealp{gir97}) and  CS (Hirano et al.\ \citealp{hir98}).  

In spite of the high interest of this region, 
there are few observations of L723 in the
optical
and near-infrared ranges. In the near-infrared, there is a K' image from Hodapp
(\citealp{hod94}), where a filamentary nebulosity is visible  $\sim
1\farcm5$ northwest
of VLA2. A more recent H$_2$ image from Palacios \& Eiroa
\cite{pal99} shows H$_2$ emission, with several condensations 
located at  both sides of VLA2, and lying projected along the 
axis direction of the east-west bipolar CO outflow.
At optical wavelengths, there are the R,\ I and
H$\alpha$ images of Vrba et al.\ \cite{vrb86}. The H$\alpha$ image shows a
nearly linear emission structure of $\sim 30''$, elongated in the east-west
direction and located towards the east of VLA2, at the southern edge
of the high extinction region  where the radio source is embedded. This
structure corresponds to HH~223 in the Reipurth Catalogue of Herbig-Haro
(HH) objects
(Reipurth \citealp{rei94}). In addition, their optical
images
show several faint nebulosities that the authors identify as  possible HH objects. 
All of them, HH~223 and
the nebulosities, lie projected onto the lobes of the east-west CO bipolar outflow.  
However,
these optical images are not deep enough to resolve the HH~223 morphology and
they were not acquired in  good weather conditions.
With the aim of studying with more detail the morphology of the emission in HH~223
and to elucidate whether the optical nebulosities have an  HH nature,
as well as to
explore the relationship between the optical, molecular and near infrared emissions
of L723, we  acquired 
new, deep optical narrow-band images in the 
H$\alpha$ and [SII] lines and in the continuum nearby H$\alpha$ of 
the L723 field. 
The results are presented in this paper.

\section{Observations and Data Reduction}

\begin{figure*}
\centering
\rotatebox{-90}{\includegraphics[width=9.25cm]{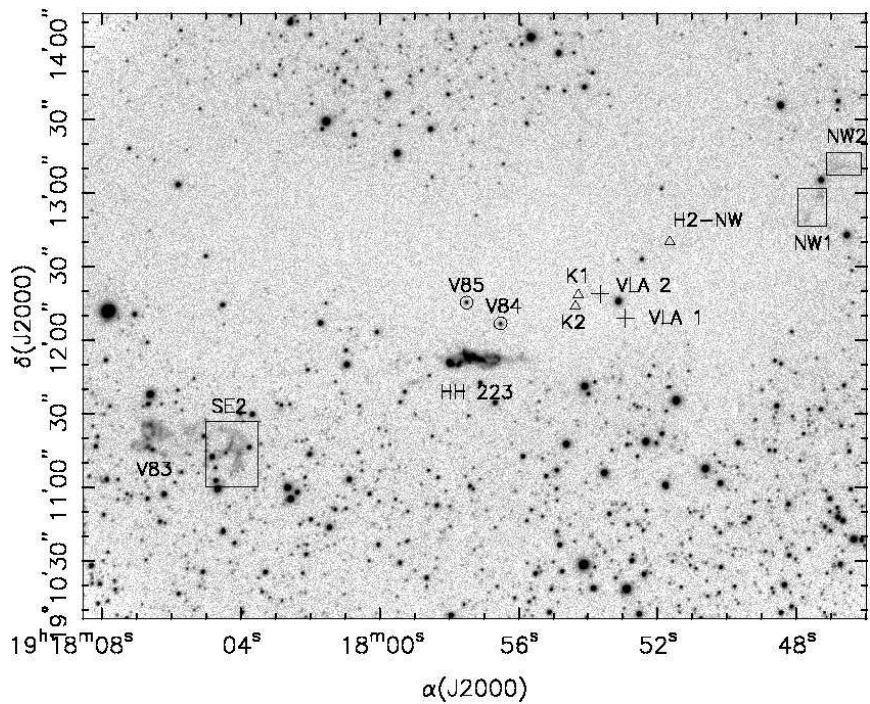}}
\caption{H$\alpha$ image of the L723 field. HH~223 is at the center of the field. 
Crosses mark the
positions of the VLA sources of Anglada et al.\  (\citealp{ang96}). The positions of 
the H$_2$ emission peaks K2, K1 and H$_2$-NW 
of Palacios \& Eiroa (\citealp{pal99}) have been marked by triangles. Open circles  
mark the stars
that coincide with the positions of the objects  V84 and V85 of 
Vrba et al.\  (\citealp{vrb86}). Rectangles mark the new H$\alpha$ nebulosities 
detected in this work. The nebulosity coinciding with the V83 object of  
Vrba et al.\  (\citealp{vrb86}) (HH~223-SE1 
in this work) has also been marked.
\label{fig1}} 
\end{figure*}

\begin{figure*}
\centering
\rotatebox{-90}{\includegraphics[width=9.25cm]{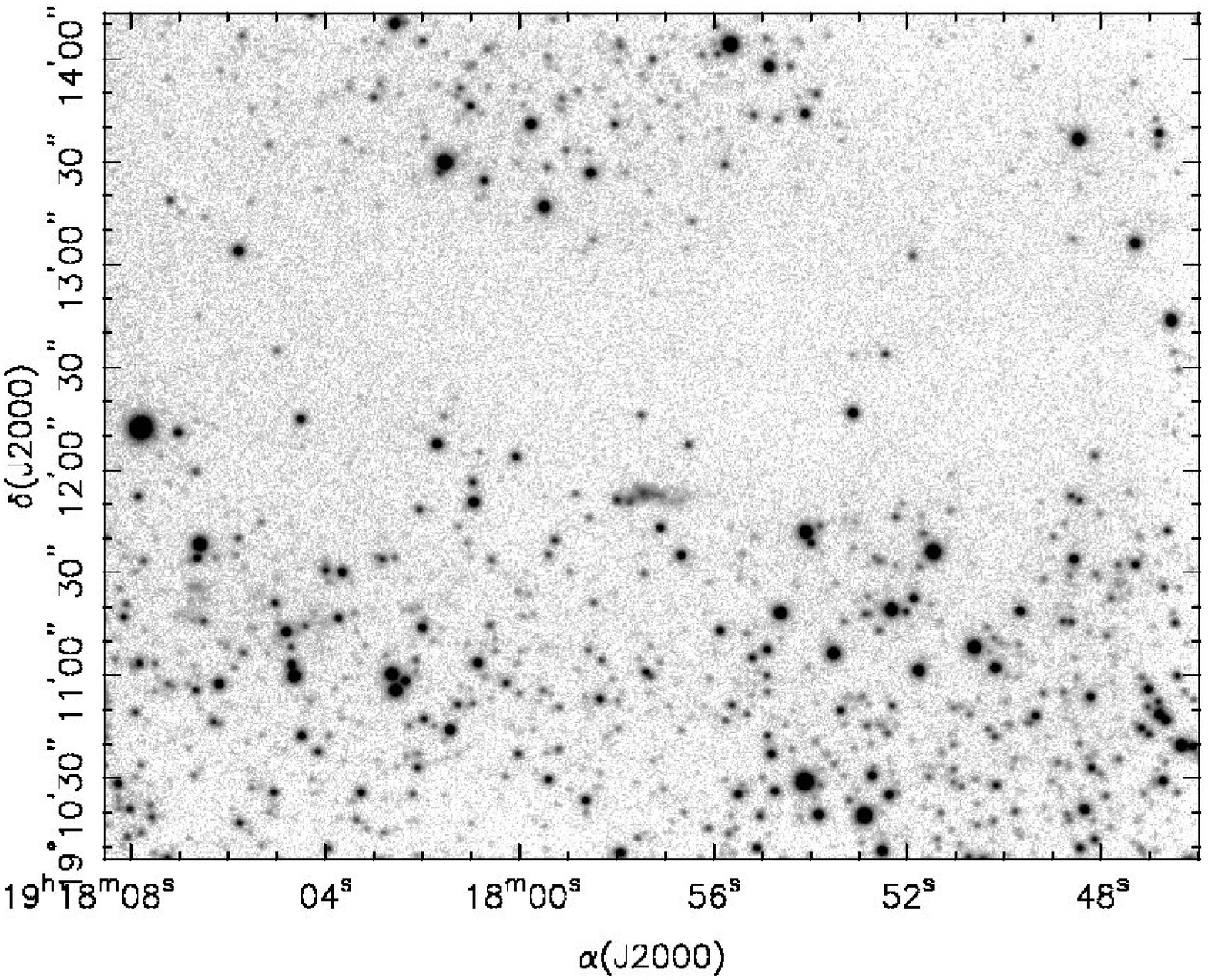}}
\caption{[SII] image of the L723 field. 
\label{fig2}} 
\end{figure*}

Deep CCD narrow-band images of L723, which includes
HH~223, were obtained 
with the 2.6~m Nordic Optical Telescope (NOT)
of the Observatorio del Roque de los Muchachos (ORM, La Palma,
Spain)  using the Service Time mode facility in two observing runs 
(July 2004 and September 2005).
The images were obtained with the Andalucia 
Faint Object Spectrograph and Camera (ALFOSC). The 
image scale was $0\farcs188$~pixel$^{-1}$. The 
effective imaged field was $\sim5'\times5'$. 
In July 2004, an H$\alpha$ filter (central wavelength
$\lambda=6564$~\AA, bandpass $\Delta\lambda=33$~\AA ) was used to obtain 
a deep image (2.5~h 
integration time) in the H$\alpha$ line by combining five frames of 
1800~s exposure each.
In addition, a filter of the continuum nearby H$\alpha$ 
($\lambda=6621$~\AA, $\Delta\lambda=39$~\AA ) was used 
to obtain an
image of the same field of 900~s exposure. 
The range of this continuum filter excludes bright HH emission lines,
being thus useful to discriminate between scattered starlight and pure
H$\alpha$ line emission. 
Values of the
seeing were $0\farcs7$--$1''$.
In September 2005, a [SII]  filter 
($\lambda=6725$~\AA, $\Delta\lambda=60$~\AA ) that includes the two red
[SII] lines, was used to obtain 
a deep image (2.5~h 
integration time) by combining five frames of 
1800~s exposure each. Values of the
seeing were $1\farcs4$--$1\farcs8$.

All the images were processed with the standard tasks of the IRAF\footnote{IRAF is
distributed by the National Optical Astronomy Observatories, which are
operated by the Association of Universities for Research in Astronomy,
Inc., under cooperative agreement with the National Science Foundation.}
reduction package. In order to correct for the misalignments among 
individual frames, the
frames were recentered using the reference
positions of fifteen field stars. 
Astrometric calibration of the three final images 
was performed in order to compare the positions of  
HH~223 and other line-emission nebula detected in the field with 
the positions of the VLA2 radio continuum source,
and with the other molecules 
(i.e., H$_2$, CO). 
In order
to register the images, we used the $(\alpha, \delta)$ coordinates 
from the  USNO-B1.0 Catalogue\footnote{The USNOFS Image and 
Catalogue Archive is operated by the United States Naval Observatory, 
Flagstaff Station.} of ten field stars well distributed on the 
observed field. The rms of the transformation was $0\farcs16$ in both coordinates.

In addition, with the aim of performing 
a meaningful  comparison between the
H$\alpha$ and [SII] emissions of HH~223, we extracted 
from the [SII] and H$\alpha$ images of the L723 field two
subimages of $85'' \times 35''$  each. 
Both subimages included HH~223 and ten field stars. 
The H$\alpha$ subimage was smoothed with a Gaussian to match the seeing of the [SII] subimage.
The background, obtained by averaging the emission of several regions 
free of stars 
close to HH~223, was substracted to each image. Finally, the subimages were scaled 
in flux by applying a scaling factor derived from
differential aperture photometry of the common field stars of the two subimages.
These two subimages were then used to obtain the HH~223 [SII]/H$\alpha$ flux
ratio map that will be discussed later.

\section{Results and Discussion}

Figure\ \ref{fig1}  shows  the H$\alpha$ line image of L723   
and Figure\ \ref{fig2} shows the [SII] image of the same field.
At the center of both images there
is a bright, extended structure of $\sim 30''$ length, elongated
approximately along the east-west 
direction that corresponds to HH~223, first reported by
Vrba et al. \  \cite{vrb86} as a ''linear emission feature''.
In addition, the  image reveals several other fainter nebulosities 
 to the south-east and to the north-west of HH~223.
All of these H$\alpha$ nebulosities are undetected in the image 
obtained through the narrow-band filter 
that includes the emission of the continuum nearby H$\alpha$, 
indicating thus that all these features are
pure emission-line features.
We will describe all of them in the following.
 
\subsection{HH~223 morphology}

\subsubsection{H$\alpha$ emission}

\begin{figure*}
\centering
\rotatebox{90}{\includegraphics*[width=7cm]{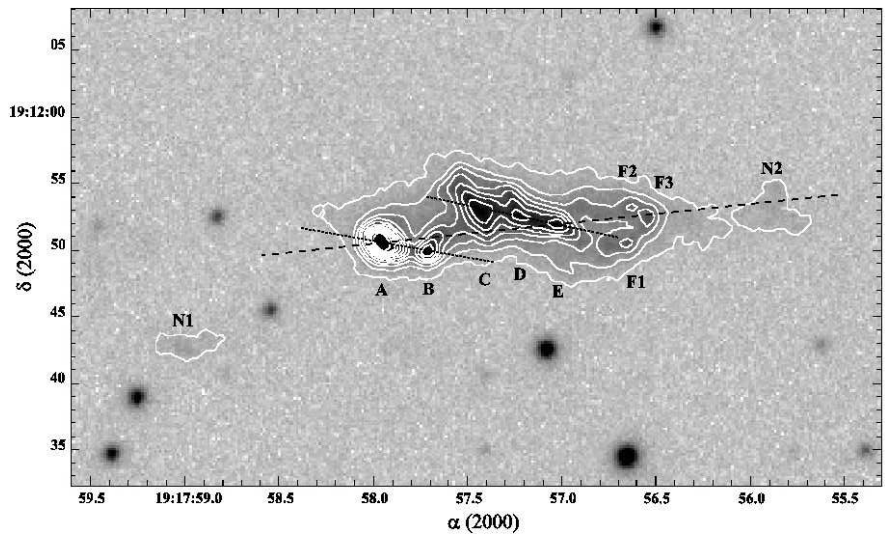}}
\caption{Close-up of HH~223 and the two adjacent nebulosities N1 and N2 obtained 
from the H$\alpha$ image of Fig.\ \ref{fig1}, showing  
the details of the emission in HH~223. 
Several knots
appear embedded in a
low-emission nebula. A contour plot  has
been superposed to help with the identification of the knots. 
Dashed and dotted lines mark the orientations refered to in the text.
\label{fig3}} 
\end{figure*}

Our H$\alpha$ image shows unprecedent details of the morphology of HH~223 
and allow us to resolve the structure of the ''linear emission feature''
reported by Vrba et al.\  \cite{vrb86}.
Our image reveals a quite complex knotty structure, with several 
condensations
embedded in a more diffuse, nebular emission. 
 Figure \ref{fig3} shows a close-up of the H$\alpha$ image of HH~223. 
We have identified several bright
knots, labeled from  A to F3, from  east to  west. 
Table \ref{tjet} gives the peak positions of the knots.
It should be noted that knots A and F2 coincide,
respectively, with the positions given by Vrba et al.\  \cite{vrb86} for the L723 
objects V81 and V82, which
they used to mark the  eastern and western borders of their ''linear emission
feature''.
Furthermore, we detected in the H$\alpha$ image two additional nebular emissions  
that do not
appear in the continuum image. 
One of these nebulosities (N2) extends over $\sim 11''$ to the west 
of the F3 knot and most probably forms part of the HH~223 object.
Another faint, arc-shaped  nebulosity (N1) extends over $\sim$ $3\farcs5$, 
and is located
$\sim 13''$ to the SE of HH~223 (see Fig.\ \ref{fig3}).

\begin{table}
\caption[ ]{Peak positions of the H$\alpha$ knots of HH~223 and H$\alpha$ nebulosities
 in L723}
\label{tjet}
\centering
\begin{tabular}{lccc}  
\hline
Object & $\alpha_{2000}$ & $\delta_{2000}$  & Notes \\
\hline 
HH~223 knots:       &              &             & \\ 
\phantom{0}A  & 19 17 57.97  &+19 11 50.7  & \\
\phantom{0}B  & 19 17 57.72  &+19 11 50.1  & \\
\phantom{0}C  & 19 17 57.43  &+19 11 53.3  & \\
\phantom{0}D  & 19 17 57.30  &+19 11 53.0  & \\
\phantom{0}E  & 19 17 57.03  &+19 11 52.1  & \\
\phantom{0}F1 & 19 17 56.66  &+19 11 50.9  & (1)\\
\phantom{0}F2 & 19 17 56.66  &+19 11 53.4  & (1)\\
\phantom{0}F3 & 19 17 56.56  &+19 11 52.9  & (1)\\  
H$\alpha$ nebulosities:&           &      & \\ 
\phantom{0}N1 & 19 17 59.0  &+19 11 42   & (2)\\  
\phantom{0}N2 & 19 17 55.0  &+19 11 53   & (2)\\
\phantom{0}HH 223-SE1& 19 18 06.1  &+19 11 21   &(2)\\
\phantom{0}HH 223-SE2& 19 18 04.3  &+19 11 16   &(2)\\
\phantom{0}HH 223-NW1& 19 17 47.3  &+19 13 01   &(2)\\
\phantom{0}HH 223-NW2& 19 17 46.8  &+19 13 12    &(2)\\
\hline
\end{tabular}
\begin{list}{}{}
\item[] Notes:\\
(1) Peak positions of the emission enhancements found in the faint extended
arc-like filamentary emission westwards of knot E.\\
(2) Position of the centroid of the nebulosity.\\
\end{list}
\end{table}

Palacios \& Eiroa \cite{pal99} detected H$_2$ emission
in L773. The emission shows 
several knots lying along the axis direction of the 
east-west CO bipolar outflow.
The H$_2$ knots labeled H2-SE by these authors 
appear to be the near infrared counterpart of HH~223. 
A more careful comparison between the  near infrared 
(H$_2$) and the optical (H$\alpha$) emissions shows that 
the H2-SE emission of 
 Palacios \& Eiroa \cite{pal99}
spatially coincides with the H$\alpha$ emission from knots C to F3. 
Both emissions  also show very similar shapes, 
although the H$\alpha$  emission is spatially more extended than
the H$_2$ emission. Furthermore a fainter, arc-shaped H$_2$ emission 
located to the north-west of H2-SE coincides in position with
the fainter H$\alpha$ emission nebulosity N2. 
Figures  \ref{fig1} and \ref{fig3} show that the strongest H$\alpha$ 
emission in HH~223
corresponds to knots A and B. It is not possible to know whether these H$\alpha$ 
knots have any near infrared counterpart, as the H$_2$ image of 
Palacios \& Eiroa \cite{pal99} does not
cover the region beyond the east border of HH~223 knot C. 
We have not found in our H$\alpha$ image
any optical counterpart for the near infrared emission  H2-NW  
as well as for 
the near infrared knots K1 and K2, found a few
arcsec to the east of the VLA2 radio source (see Fig.~\ref{fig1}). 
The lack of detection of optical counterparts 
of these near infrared emissions is most probably due to the higher 
visual extinction. 
However, it is interesting to point out that we have
barely detected diffuse H$\alpha$ nebulosities farther away 
($\sim$ $1\farcm5$) to the west of the VLA
radio sources. We have labeled these features HH 223-NW1 and 
HH 223-NW2 (see Table \ref{tjet} and Fig.~\ref{fig1}). 
These faint 
H$\alpha$ emissions features are found nearly aligned with the H$_2$ 
condensations K1, K2 and  H2-NW of Palacios \& Eiroa \cite{pal99}.

\begin{figure*}
\centering
\rotatebox{-90}{\includegraphics[width=5.0cm]{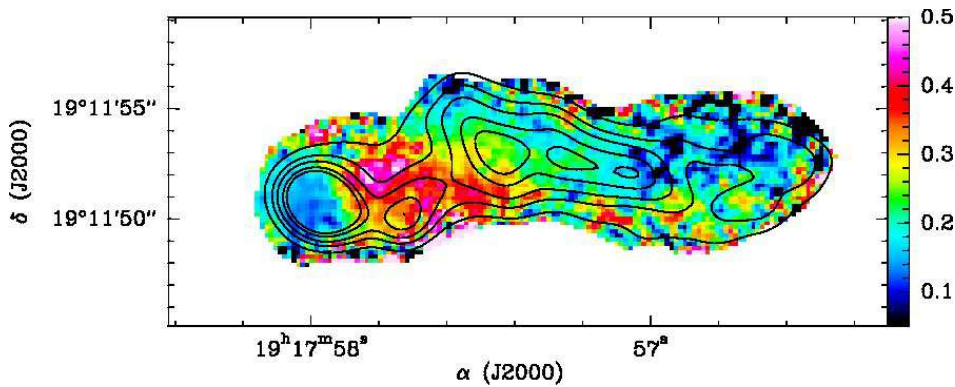}}
\caption{[SII]/H$\alpha$ flux ratio for HH~223. A contour plot of the  H$\alpha$ line
flux has been superposed.
\label{fig4}} 
\end{figure*}

As it was first pointed out by Vrba et al.\  \cite{vrb86}, 
HH~223 has a nearly east-west orientation. 
Let us define the HH~223 outflow axis passing through knots 
A-E-F3 and through the nebulosity N2 (Fig. \ref{fig3}). 
This direction defines the overall orientation of 
HH~223, taking into account the fainter H$\alpha$  nebular emission 
in which the knotty structure is found  engulfed. 
The emission axis defined in this way has a P.A. $\simeq 96^\circ$. 
This orientation is very close
to the magnetic field direction measured by Vrba et al.\  \cite{vrb86} 
(P.A.$=90^\circ$ $\pm$ 8$^\circ$) and to the axis of the 
east-west CO outflow  (P.A.$=110^\circ$).
But in addition, our deeper H$\alpha$  image allows to distinguish 
some "wiggling" in 
the HH~223 knot pattern: knots C-D-E-F1 are oriented following a 
direction with a P.A. of 79$^\circ$. 
Nearly the same orientation is obtained
for the knots A-B direction (P.A$=80^\circ$). 
Thus, despite the H$\alpha$ low-emission  presents a nearly east-west 
orientation, the knotty struture embedded 
inside the nebula is organized in two
nearly parallel chains of knots, being these chains oriented in a 
slightly different direction
within the low-emission nebula.

\subsubsection{[SII] emission}

HH~223 also shows in the [SII] lines the knotty, wiggling
structure found in H$\alpha$ (Fig.~\ref{fig2}): 
knots from A to D are
clearly identified, and [SII] emission is  detected 
at the positions found for  knot E and filament F in the H$\alpha$ image. 
However, it is not possible to 
resolve in the [SII] image 
the three condensations (F1, F2 and F3) identified  within the F filament in 
H$\alpha$.  
Furthermore, we have not detected [SII] emission coinciding
with the positions of the weaker H$\alpha$ emission nebulosities 
N1 and N2.

Interestingly, we can appreciate significant differences by comparing 
the flux  variations from knot to knot in H$\alpha$ and in [SII].
This fact is more evident for knot A, where the H$\alpha$ emission is 
stronger as compared with the rest of the knots, while  
the strength of the [SII] emission in knots A, B and C is comparable. 
For the case of  shock-excited emission  found in stellar jets, 
information on the degree of excitation of the emitting gas and on 
the shock velocity can be derived from the [SII]/ H$\alpha$ line ratio 
(Raga, B\"ohm \& Cant\'o  \citealp{rag96}; 
Hartigan, Raymond \& Hartmann \citealp{har87}). 
Thus, our data suggest variations of  
the shock-excitation conditions through HH~223.

We show 
the map (Fig.~\ref{fig4}) of the [SII] to H$\alpha$ flux ratio through HH~223 
derived from the subimages mentioned in \S 2.
We do not intend to compare the numerical values found for the [SII]/H$\alpha$ ratio 
with the line ratio values obtained from modeling the spectra of shock-excited gas.
However, our HH~223 [SII]/ H$\alpha$ map is useful to show clear signs of
variation of the excitation conditions through HH~223, as 
the [SII]/ H$\alpha$ flux ratio changes up to a
factor $\ga$ 3 in different HH~223 knots. As a general
trend, the [SII]/ H$\alpha$ flux ratio increases (excitation decreases)
from east to west 
and from north to south through
HH~223. This ratio reaches 
a minimum around knot A  (the lowest line ratio values 
are found to the south-east, close to the peak of knot A). The ratio 
has a steep increase going to the west, reaching the highest values  
towards the north of the peak of knot B. From the western edge of knot 
B to knot C 
the ratio  decreases. Beyond  knot C, the ratio remains nearly constant, 
with a slight decrease as we move from east to west and from south to north.
We especulate
that inhomogeneities, both in the ambient and in the high velocity molecular gas might be
contributing to the variations in the excitation conditions through 
HH~223 
The existence of such inhomogeneities is expected since 
HH~223 is located at an interface, 
towards the southern edge of a high
optical extinction region
and, in addition, HH~223 lies
coinciding in position with high velocity molecular gas, 
traced by CO emission, as  can be seen from Fig.~\ref{fig5}.

\subsection{Other H$\alpha$ nebulosities in L723}

\begin{figure*}
\centering
\includegraphics*[width=16cm]{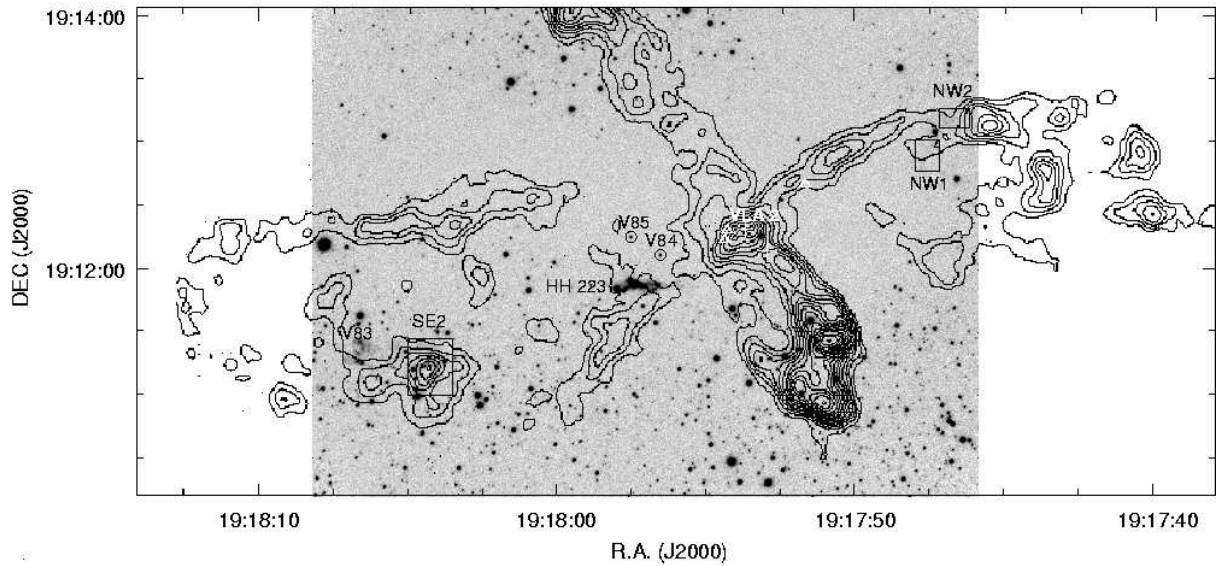}
\caption{Image of L723 in H$\alpha$ with HH~223 in the center of the field,
superposed on 
the CO outflows from Lee et al.\ \cite{lee02}. The VLA2 source 
and the other H$\alpha$ nebulosities have been labeled. The triangles mark the positions of the H$_2$ 
emission peaks of Palacios \& Eiroa \cite{pal99}.
\label{fig5}} 
\end{figure*}

We have identified some other nebulosities in the H$\alpha$ 
image of the L723 field (see Fig.~\ref{fig1} and Table \ref{tjet}
for identification).
The two brighter nebulosities are found
to the south-east of HH~223. The eastern one, $\sim 2'$ from HH~223 knot A,
coincides in position with 
the  object V83 of Vrba et al.\  \cite{vrb86}, 
although the line emission appears more extended in our image
and because of that, we have renamed it HH~223-SE1.
This nebulosity shows an
arc-shaped morphology, with its apex pointing toward HH~223 and probably has a knotty
structure which cannot be well resolved from our image. The other nebulosity, labeled  
HH~223-SE2, is $\sim 30''$ from HH~223-SE1, and 
appears as a quite bright
linear filament  $\sim 2''$ long emerging from a faint star.
Two more nebulosities are barely detected on the  H$\alpha$ image and are found 
$\sim 3'$ to the north-west of HH~223 knot A. These two features are
labeled  HH~223-NW1 and HH~223-NW2.  
Interestingly, HH~223-NW2 spatially coincides with a filamentary 
emission nebula visible on the K' image of Hodapp
(\citealp{hod94}). 
It is not possible to say from these data whether all these nebulosities 
have any relationship 
with the molecular outflows or with the near infrared and optical
line emissions found in  L723. 
However, as  can be seen in Fig. \ref{fig5}, 
all of these H$\alpha$ nebulosities are
found nearly aligned along the direction defined by the axis of the east-west bipolar
CO outflow.
In contrast, the positions
given by Vrba et al.\  \cite{vrb86} for the possible HH objects  
V84 and V85 
coincide in all of our images  with the positions of 
two faint field stars. Furthermore, these two objects are found in the 
2MASS All Sky Catalog 
of Point Sources
as two stars with K magnitudes of 10.84 and 9.38, respectively. Thus 
we discard the HH-like nature of Vrba objects V84 and V85.

\section{Conclusions}

We have obtained deep optical images in the 
H$\alpha$ and [SII] lines, and in the continuum nearby H$\alpha$  
of L723.
The H$\alpha$ image allows to resolve 
the detailed structure of the ''linear emission feature'' HH~223,
reported by Vrba et al.\  \cite{vrb86}. 
The emission consists of several knotty condensations
engulfed in  a fainter nebula.
Some wiggling is present in the HH~223 knot pattern. 
The H$\alpha$ emission spatially
coincides with an H$_2$ elongated emission, although the H$\alpha$ 
emission appears more extended.
HH~223 is also detected in the [SII] lines and shows 
the knotty, wiggling morphology found  
in the H$\alpha$ line, although the [SII] emission is weaker than the 
H$\alpha$ emission.  
In contrast, no emission was detected in the continuum image at the position of 
HH~223. Thus, HH~223 appears as a line-emission object whose 
properties are reminiscent of those
found in stellar jets, indicating a shock-excited origin of the 
emission.
Comparison between the [SII] and H$\alpha$ emission fluxes  indicates 
that there are differences in the excitation conditions through HH~223.
As a general trend, the [SII]/H$\alpha$ line ratio  increases (excitation decreases) 
from east to west and from north to south of HH~223.  
The lowest ratio (highest excitation) is found around
knot A, while the highest ratio (lowest excitation) is 
found in a region to the north of the peak of knot B. 
High resolution spectroscopy of HH~223 would be very helpful 
to confirm 
the shock-excited origin of the HH~223 emission as well as to measure 
the HH~223 radial velocity field. Knowledge on the kinematics of HH~223
would allow to establish whether the HH~223 line emission is supersonic 
(as in typical stellar jets) or, in contrast, is nearly stationary as 
would be expected for an illuminated wall of the cavity excavated by the 
molecular outflow.
 
In addition to HH~223, we have detected in the  L723
field imaged emission in the  H$\alpha$ line from several other  
nebulosities. 
None of these nebulosities
appear in the image obtained in the continuum nearby H$\alpha$. 
All of these nebulosities are found projected onto the east-west
pair of lobes of the quadrupolar L723 CO outflow. The eastern nebulosity 
coincides with the Vrba object V83, thus confirming the HH-like nature of it. 
In contrast, we 
have discarded the Vrba objects V84 and V85 as being HH-like objects.

\begin{acknowledgements} 
We acknowledge the Support Astronomer Team of the IAC for obtaining the images.
The work of R.E., R.L. and A.R. was supported by the Spanish MEC grant AYA2005-08523-C03-01.
ALFOSC is owned by the Instituto de Astrof\'{\i}sica de Andaluc\'{\i}a (IAA) and operated at the
NOT under agreement between IAA and the NBIfAFG of the Astronomical Observatory of Copenhagen.
This publication makes use of data products from the Two Micron All Sky
Survey, which is a joint project of the University of Massachusetts and the
Infrared Processing and Analysis Center/California Institute of Technology,
funded by the National Aeronautics and Space Administration and the National
Science Foundation.
\end{acknowledgements}

\end{document}